\documentclass[twocolumn]{article}
\usepackage{colortbl}
\usepackage[table]{xcolor}

\usepackage{multirow}
\usepackage{makecell}
\usepackage[normalem]{ulem}
\useunder{\uline}{\ul}{}
\usepackage{array}
\usepackage{amsmath}
\usepackage{stfloats}
\usepackage{graphicx}
\usepackage{amsfonts}
\usepackage{mathtools}
\usepackage{booktabs}
\usepackage{listings}
\usepackage{url}
\usepackage{csquotes}
\usepackage{enumitem}
\setlist[itemize]{nosep}
\usepackage{authblk}
\usepackage{caption}
\usepackage[a4paper, left=0.68in, right=0.68in, top=0.75in, bottom=1.0in]{geometry}
\makeatletter
\renewcommand{\section}{\@startsection{section}{1}{\z@}{1.5ex plus .5ex minus .2ex}%
{1.3ex plus .2ex}{\centering\normalfont\normalsize\bfseries}}
\makeatother
\makeatletter
\renewcommand{\subsection}{\@startsection{subsection}{2}{\z@}{1.5ex plus .5ex minus .2ex}%
{1.3ex plus .2ex}{\centering\normalfont\normalsize\bfseries}}
\makeatother

\newcommand{\codesty}[1]{\texttt{#1}}
\definecolor{keywordcolor}{HTML}{385fff} 
\definecolor{commentcolor}{HTML}{41ab2c}
\definecolor{stringcolor}{HTML}{ff9107} 
\author[1,2]{Chengxuan Qin \thanks{C.Qin8@liverpool.ac.uk}}
\author[1]{Rui Yang \thanks{Corresponding author: R.Yang@xjtlu.edu.cn}}  
\author[1,2]{Wenlong You}
\author[3]{Zhige Chen}
\author[1,2]{Longsheng Zhu}
\author[4]{Mengjie Huang} 
\author[5]{Zidong Wang} 

\affil[1]{School of Advanced Technology, Xi'an Jiaotong-Liverpool University, Suzhou, 215123, China}
\affil[2]{School of Electrical Engineering, Electronics and Computer Science, University of Liverpool, Liverpool, L69 3BX, United Kingdom}
\affil[3]{Department of Computing, The Hong Kong Polytechnic University, Hong Kong, SAR}
\affil[4]{Design School, Xi'an Jiaotong-Liverpool University, Suzhou, 215123, China}
\affil[5]{Department of Computer Science, Brunel University London, Uxbridge, Middlesex, UB8 3PH, United Kingdom}
\begin{document}

\title{EEGUnity: Open-Source Tool in Facilitating Unified EEG Datasets Towards Large-Scale EEG Model}

\maketitle

\begin{abstract}
The increasing number of dispersed EEG dataset publications and the advancement of large-scale Electroencephalogram (EEG) models have increased the demand for practical tools to manage diverse EEG datasets. However, the inherent complexity of EEG data, characterized by variability in content data, metadata, and data formats, poses challenges for integrating multiple datasets and conducting large-scale EEG model research. To tackle the challenges, this paper introduces EEGUnity, an open-source tool that incorporates modules of \enquote{EEG Parser}, \enquote{Correction}, \enquote{Batch Processing}, and \enquote{Large Language Model Boost}. Leveraging the functionality of such modules, EEGUnity facilitates the efficient management of multiple EEG datasets, such as intelligent data structure inference, data cleaning, and data unification. In addition, the capabilities of EEGUnity ensure high data quality and consistency, providing a reliable foundation for large-scale EEG data research. EEGUnity is evaluated across 25 EEG datasets from different sources, offering several typical batch processing workflows. The results demonstrate the high performance and flexibility of EEGUnity in parsing and data processing. The project code is publicly available at github.com/Baizhige/EEGUnity.
\end{abstract}


\footnotetext[1]{This research has been approved by University Ethics Committee of Xi’an Jiaotong-Liverpool University with proposal number EXT20-01-07 on March 31 2020, and is partially supported by: National Natural Science Foundation of China (72401233), Jiangsu Provincial Qinglan Project, Natural Science Foundation of the Jiangsu Higher Education Institutions of China (23KJB520038), and Research Enhancement Fund of XJTLU (REF-23-01-008).}

\section{Introduction}

Brain-computer interfaces (BCI) are really pushing the limits, being able to record over not just days, weeks, but months, years at a time \cite{commentBCI_Naddaf_2024}. The proposal of BCI systems has encouraged many researchers to actively explore brain signals and apply BCI systems in various fields: game interaction entertainment, robot control, emotion recognition, fatigue detection, sleep quality assessment, and clinical fields \cite{robot_control_Berdell_2024, BCIReview_Gu_2021, BCIReview_Naser_2023, EmoReview_Somarathna_2023, NVAR_Wang_2023, MIReview_Wang_2023}. With the increasing exploration of BCI systems, the demand for neuro-monitoring capabilities has significantly increased. EEG is applied in various BCI systems to collect active electrical signals from the brain. To meet diverse requirements, numerous EEG-based BCI system paradigms have been proposed \cite{ReviewParadigm_Abiri_2019}. The paradigms allow for the selection of an appropriate approach to gather extensive brain electrical signals to form EEG datasets that provide foundational knowledge of specific patterns in the brain \cite{pattern_Shivaprasad_2024, DifferentBrainPatternsU0_2023_guan}.

In recent years, there has been a marked increase in both the quantity and demand for EEG data publications. On the one hand, online databases such as Zenodo \cite{database_zenodo} and PhysioNet \cite{database_PhysioNet} have a vast and growing collection of diverse EEG datasets. On the other hand, a study \cite{LbraM_Jiang_2024} on large EEG models employing over 20 datasets demonstrates the ever-increasing need to process large-scale EEG data. Nevertheless, the inherent characteristics of EEG data present a challenge in data processing \cite{noisy_data_Khan_2023, ChallengesDataProcessingU0_2023_zhou}. The challenge is primarily attributed to three key factors:
\begin{itemize}
    \item \textbf{Differences in content data}: Variations in the configuration of electrodes, characteristics of the sensor, and circuit structures can lead to significant differences in the dimensions and distribution of EEG data \cite{EEGDevcieReview_Guiomar_2023, ChannelSel_Wei_2023, LTA_Yi_2023};
    \item \textbf{Differences in metadata}: Variations in the labeling criteria for metadata (such as channel names and events annotation), along with the absence or errors in standard information, can lead to inconsistencies in annotation \cite{MNEPython_Alexandre_2013};
    \item \textbf{Differences in data formats}: Variations in data formats (such as gdf, edf, mat, csv, txt) complicate the standardization and processing steps of data, demonstrating that the processing pipeline must be tailored for each study \cite{MNEPython_Alexandre_2013, FieldTrip_Robert_2011}.
\end{itemize}

\begin{table}[b]
\fontsize{8}{10}\selectfont
\captionsetup{font={footnotesize}}
\centering
\caption{Overview of Commonly Used EEG Data Processing Software. \label{tab_overview_eeg_softwares}}
\begin{tabular}{llll}
\rowcolor[HTML]{F2F2F2} 
\hline
\textbf{Tool} & \textbf{Openness} & \textbf{Platform} \\ \hline \hline
SPM  \cite{SPM_Vladimir_2011}             & open source   & MATLAB \\  
EEGLAB \cite{EEGLAB_Delorme_2004}            & open source   & MATLAB \\
FieldTrip \cite{FieldTrip_Robert_2011}         & open source   & MATLAB \\
Brainstorm \cite{Brainstorm_François_2011}       & open source   & MATLAB \\
MNE-Python \cite{MNEPython_Alexandre_2013}       & open source   & Python \\ 
NeuroScan CURRY \cite{NeurScanCurryUsed_Darion_2022}  & commercial    & Windows \\
BESA \cite{BESAIntro_Aya_2023}             & commercial    & Windows \\ \hline
\end{tabular}
\end{table}

The challenges above indicate an urgent need for an efficient large-scale EEG data management tool to simplify data standardization and processing steps, eventually improving data processing efficiency. Table \ref{tab_overview_eeg_softwares} summarizes highly integrated EEG data processing tools, covering their openness and supported platforms. Various EEG data processing tools with comprehensive functionality are widely used in multiple EEG research. Some of the common open-source tools include SPM \cite{SPM_Vladimir_2011}, EEGLAB \cite{EEGLAB_Delorme_2004}, FieldTrip \cite{FieldTrip_Robert_2011}, Brainstorm \cite{Brainstorm_François_2011}, and MNE-Python \cite{MNEPython_Alexandre_2013}, offering flexible and powerful analysis capabilities to meet diverse research needs. Additionally, commercial software such as NeuroScan CURRY \cite{NeurScanCurryUsed_Darion_2022} and BESA \cite{BESAIntro_Aya_2023} are also available.

Despite the significant advantages in functionality and broad applicability offered by existing tools, the tools listed in Table \ref{tab_overview_eeg_softwares} exhibit shortcomings when handling large-scale data. Specifically, the existing tools offer flexible processing for individual datasets, but lack specialized management methods to simultaneously handle multiple datasets with varying content data, metadata, and formats. Therefore, current tools are limited in effectively managing and analyzing heterogeneous EEG data from various sources. As the quantity and demand for EEG datasets continue to increase, the shortcomings become particularly evident, affecting research efficiency and the reliability of results. Therefore, an urgent need is to enhance EEG data processing capability for handling large-scale EEG data.

\begin{figure}[htbp]
    \centering
    \includegraphics[width=0.8\linewidth]{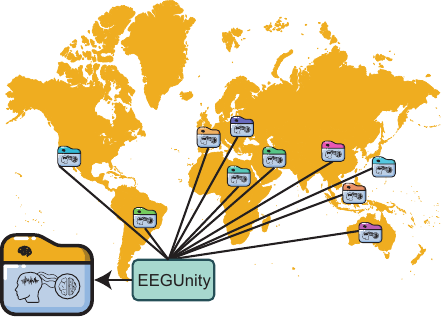}
    \caption{Objective of Proposed EEGUnity. \label{fig_vision}}
\end{figure}

In this paper, a novel EEG data processing tool named EEGUnity is proposed, aiming to unify diverse EEG datasets from all over the world, as reflected in its name. As illustrated in Fig. \ref{fig_vision}, the objective of EEGUnity is to manage multiple datasets, efficiently process large-scale data, and enhance data processing efficiency. The introduction of EEGUnity addresses the limitations of existing tools in handling large-scale EEG data and provides a new approach to the unified management and processing of EEG data. EEGUnity offers several innovative features: (1) intelligent data structure inference technology to address the challenge of data heterogeneity; (2) a user-friendly interface for reviewing and modifying EEG dataset annotations to ensure accurate analysis; (3) a comprehensive and unified interface for large-scale data processing to establish a solid foundation for subsequent analyses. These features allow EEGUnity to effectively manage and process heterogeneous EEG data from different sources, enhancing data consistency and comparability.

Based on the above discussions, the main contributions of this paper can be summarized as follows: 
\begin{itemize}
    \item The proposal of a new concept for efficiently managing and processing large-scale EEG data through a unified platform;
    \item The proposal of EEGUnity, a tool specifically designed for EEG data processing, supporting the management of multiple datasets, thereby addressing the challenges of data heterogeneity;
    \item The intelligent integration of features into EEG Unity, including data structure inference, correction, cleaning, and unification, ensuring high data quality and consistency, thereby providing a reliable data foundation for EEG research.
\end{itemize}

The remaining structure of this paper is arranged as follows: Section \ref{section_method} provides a detailed introduction to EEGUnity, including overview and implementation details, giving readers comprehensive background information about the tool. Section \ref{section_experiment} demonstrates several typical batch process workflows of EEGUnity, such as dataset management, data correction, data cleaning, and data unification, through specific use cases, helping readers understand the practical application of the tool. Section \ref{section_discussion} comprehensively analyzes the advantages and limitations of EEGUnity in practical applications, and Section \ref{section_conclusion} concludes this paper and proposes future research directions.

\section{Introducing EEGUnity: Overview and Implementation \label{section_method}}

In this section, an overview of EEGUnity is initially presented, including the implementation platform, components, usage, and innovative features. Subsequently, the architectural design of two core components of EEGUnity, \codesty{UnifiedDataset} and \codesty{Locator}, is detailed.

\subsection{Overview of EEGUnity}
\begin{figure*}[htbp]
    \centering
    \includegraphics[width=0.80\linewidth]{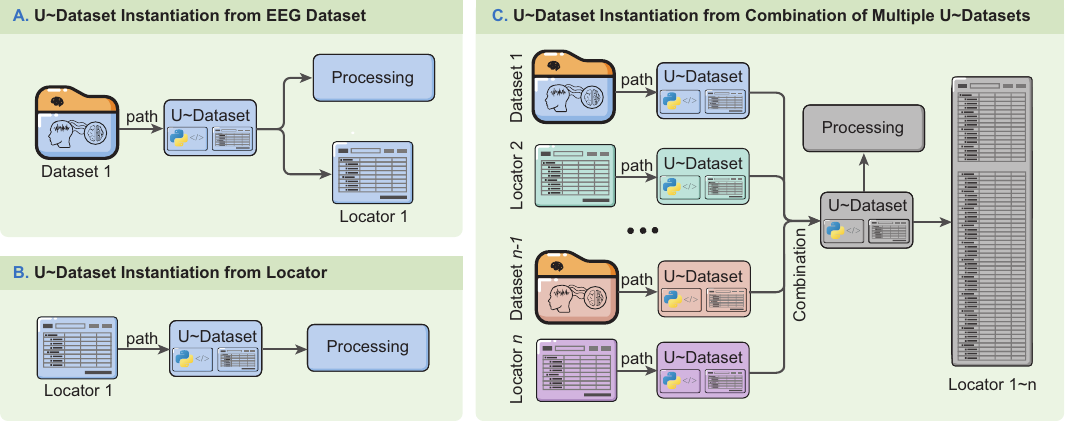}
    \caption{Three Approaches for Managing Datasets in EEGUnity. \enquote{U$\sim$Dataset} refers to \codesty{UnifiedDataset}. \label{fig_workflows_eegunity}}
\end{figure*}
EEGUnity is currently a Python package focused on managing multiple EEG datasets. The EEGUnity package includes a core Python class---\codesty{UnifiedDataset}, along with functions for creating, modifying, and merging on the \codesty{UnifiedDataset}. The Python class \codesty{UnifiedDataset} is designed to provide a unified interface for different operations to datasets. Users can efficiently manage multiple datasets through \codesty{UnifiedDataset}, thereby gaining the capability to handle large-scale EEG data.

The usage for managing datasets in EEGUnity is very convenient, following a two-step process: 1) instantiating a \codesty{UnifiedDataset}; 2) performing batch processing based on the \codesty{UnifiedDataset}. There are three approaches for managing datasets in EEGUnity, as shown in Fig. \ref{fig_workflows_eegunity}:
\begin{itemize}
    \item The first approach is illustrated in Fig. \ref{fig_workflows_eegunity}A. EEGUnity supports the instantiation of \codesty{UnifiedDataset} by providing an accessible path to EEG dataset for the initial accessing. After instantiating the \codesty{UnifiedDataset}, users can utilize the interface of \codesty{UnifiedDataset} for batch processing or export a \codesty{Locator}, which is a file recording the essential metadata that required to access the dataset, while parser processes are automatically employed by \codesty{UnifiedDataset}.
    \item The second approach is illustrated in Fig. \ref{fig_workflows_eegunity}B. EEGUnity supports the instantiation of \codesty{UnifiedDataset} by utilizing a pre-existing \codesty{Locator}. If the dataset referenced by the \codesty{Locator} remains accessible on the user's system, the resulting \codesty{UnifiedDataset} instance is equivalent to one that would have been instantiated directly via an EEG dataset address. This design allows users to quickly reload a dataset and save any modification made to the metadata.
    \item The third approach is illustrated in Fig. \ref{fig_workflows_eegunity}C. EEGUnity supports the instantiation of \codesty{UnifiedDataset} by integrating multiple \codesty{UnifiedDataset} instances, thereby allowing users to effectively manage multiple datasets according to specific requirements.
\end{itemize}

\begin{figure*}[bp]
    \captionsetup{font={footnotesize}}
    \centering
    \includegraphics[width=0.80\linewidth]{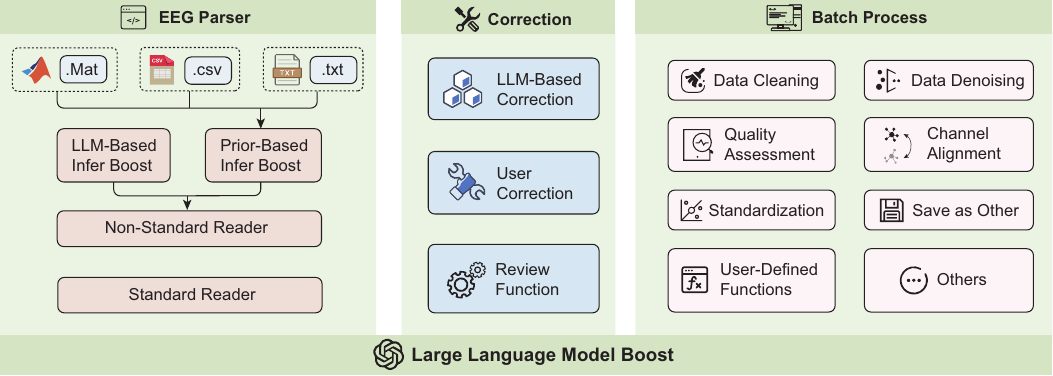}
    \caption{Structure of the UnifiedDataset. \label{fig_unifieddataset_structure}}
\end{figure*}

Compared to other processing tools, EEGUnity offers the following innovative features:
\begin{itemize}
    \item \textbf{Data Scale Capabilities}: EEGUnity is designed to process large-scale EEG datasets, supporting the integration of multiple datasets and large-scale data operations;
    \item \textbf{Functional Scope}: EEGUnity processes diverse functions for data cleaning, data correction, data unification, and custom batch processing operations, significantly improving EEG data processing efficiency;
    \item \textbf{Compatibility and Flexibility}: EEGUnity supports intelligent data structure inference using preset code and large language models (LLM), enhancing the flexibility and accuracy of data processing.
\end{itemize}

\subsection{Implementation Details of UnifiedDataset}

In EEGUnity, the parser for EEG data and the unified interface are integrated within a Python class \codesty{UnifiedDataset}. \codesty{UnifiedDataset} includes a comprehensive set of methods and attributes to provide a unified interface, as shown in Fig. \ref{fig_unifieddataset_structure}. The functions can be categorized into four modules:
\begin{itemize}
    \item \textbf{EEG parser module}: This module facilitates the intelligent parsing of EEG data across diverse formats, including but not limited to edf, gdf, mat, txt, and csv. The EEG data files are initially processed using standard readers, such as parser functions in MNE-Python. If standard readers are insufficient, non-standard readers are employed, which are boosted by pre-defined code and another module in EEGUnity---large language model boost module;
    
    \item \textbf{Correction module}: This module offers a user-friendly interface to facilitate the correction of dataset annotations. On the one hand, the module allows users to visually inspect and modify annotations within a spreadsheet-like environment. On the other hand, the module provides several correction methods, by built-in functions and the large language model boost module, to systematically review and refine dataset annotations.
    
    \item \textbf{Batch processing interface module}: This module provides batch processing functionality. On the one hand, this module enables users to customize batch processing pipelines according to specific needs. On the other hand, this module includes a variety of built-in functions for EEG data processing, such as data cleaning, denoising, quality assessment, channel alignment, and standardization.
    
    \item \textbf{Large language model boost module}: This module enhances the capabilities of other modules within \codesty{UnifiedDataset} by leveraging existing large language models, such as ChatGPT. This module enables advanced functionalities, including intelligent data parsing and the extraction of channel names and sampling rates from given descriptive files.
\end{itemize}

\subsection{Implementation Details of Locator}
In the design of EEGUnity, a critical component supporting the functionality of EEGUnity is the \codesty{Locator}. During the instantiation of \codesty{UnifiedDataset}, attributes for each EEG data file are stored in the \codesty{Locator}, which is structured using a \codesty{DataFrame} from the Pandas package---a widely adopted Python library for data science applications \cite{pandas_wes_2010}. The attributes within the \codesty{Locator} are categorized into basic and advanced attributes for each data file:
\begin{itemize}
    \item \textbf{Basic attributes} include file path, domain tag, file type, EEG channel configuration, sampling rate, duration, and completeness check;
    \item \textbf{Advanced attributes} are linked to specific submodules for functionalities such as data quality scoring. Advanced attributes can be added by subsequent community developers or customized according to user requirements.
\end{itemize}

Based on the design of \codesty{Locator}, users can calibrate the dataset by visually inspecting and modifying the \codesty{Locator} file in a spreadsheet-like environment, either programmatically or manually. The metadata specified by the \codesty{Locator} takes precedence over those in source data. In such a design, EEGUnity allows users to quickly correct metadata through the locator without modifying the source files.

\section{Typical Batch Processing Workflow in EEGUnity \label{section_experiment}}

This section introduces typical batch processing workflows in EEG data processing, including dataset management, data correction, data cleaning, and data unification. The described typical batch processing workflows comprise multiple functions of EEGUnity and are not fixed pipelines. Users can customize the pipelines according to specific requirements. Table \ref{tab_datasets} lists the datasets used in this study.
\begin{table}[htbp]
\fontsize{8}{10}\selectfont
\captionsetup{font={footnotesize}}
\centering
\caption{Datasets Used in This Study. \label{tab_datasets}}
\resizebox{\linewidth}{!}{%
\begin{tabular}{l|l|l}
\rowcolor[HTML]{E7E6E6} \hline File Name & Paradigms                                                                    & Total Hours \\ \hline \hline
zenodo-saa \cite{dataset_zenodo_saa}                               & audio attention                                                              & 15          \\ 
tuh-eeg-seizure \cite{dataset_thu_eeg}                          & annotations about seizures                                                   & 1473.6      \\
tuh-eeg-events \cite{dataset_thu_eeg}                           & annotations about epilepsy                                                   & 148.7       \\
tuh-eeg-artifact \cite{dataset_thu_eeg}                         & artifacts annotations                                                        & 100         \\
tuh-eeg-abnormal \cite{dataset_thu_eeg}                         & normal or abnormal                                                           & 1137.3      \\
tuh-eeg \cite{dataset_thu_eeg}                                  & clinical EEG recordings                                                      & 27065       \\
physionet-sleepedfx \cite{dataset_physionet_sleepedfx}                      & sleep patterns                                                               & 3849.1      \\
physionet-eegmmidb \cite{dataset_physionet_eegmmidb}                       & motor imagery                                                                & 48.5        \\
physionet-eegmat \cite{dataset_physionet_eegmat}                         & mental arithmetic                                                            & 2.4         \\
physionet-chbmit \cite{dataset_physionet_chbmit}                         & intractable seizures                                                         & 1061        \\
physionet-capslpdb \cite{dataset_physionet_capslpdb}                       & cyclic alternating pattern                                                   & 1004        \\
other-seed \cite{dataset_other_seed}                               & various                                            & 172.3       \\
other-openbmi \cite{dataset_other_openbmi}                            & various                                                                                   & 43.0        \\
other-migrainedb \cite{dataset_other_migrainedb}                         & migraine                                                                     & 21.2        \\
other-highgammadataset \cite{dataset_other_highgammadataset}                   & motor imagery                                                                & 28.7        \\
openneuro-ds004015 \cite{dataset_openneuro_ds004015}                       & audio attention                                                              & 47.3        \\
openneuro-ds003516 \cite{dataset_openneuro_ds003516}                       & audio attention                                                              & 22.6        \\
figshare-stroke \cite{dataset_figshare_stroke}                          & motor imagery                                                                & 4.4         \\
figshare-shudb \cite{dataset_figshare_shudb}                           & motor imagery                                                                & 24.8        \\
figshare-meng2019 \cite{dataset_figshare_meng2019}                        & motor imagery                                                                & 49.6        \\
figshare-largemi \cite{dataset_figshare_largemi}                         & motor imagery                                                                & 70.2        \\
bcic-iv-2b \cite{dataset_bcic_iv}                               & motor imagery                                                                & 26.3        \\
bcic-iv-2a \cite{dataset_bcic_iv}                              & motor imagery                                                                & 13.4        \\
bcic-iv-1 \cite{dataset_bcic_iv}                               & motor imagery                                                                & 3.7         \\
bcic-iii-1 \cite{dataset_bcic_iii}                               & motor imagery                                                                & 0.3         \\
\hline
\end{tabular}
}
\end{table}

\subsection{Datasets Management}
EEGUnity offers three primary approaches when initiating the processing of one or multiple EEG datasets: (1) specifying the path to an available dataset; (2) specifying the path to an available locator; (3) integrating multiple datasets. The detailed implementation for each approach is provided in Table \ref{tab_code_manage}.

\begin{table}[htbp]
\captionsetup{font={footnotesize}}
\fontsize{8}{10}\selectfont
\centering
\scriptsize
\caption{Implementation of Three Approaches for Managing Datasets in EEGUnity. \label{tab_code_manage}}

\rule{0.48\textwidth}{0.4pt}
\vspace{-\baselineskip}
\begin{lstlisting}[language=Python, basicstyle=\ttfamily, breaklines=true]
# Dataset Management Implementation in EEGUnity
# -----------------------------------------------
# This script demonstrates three approaches to instantiate and manage EEG datasets using EEGUnity toolkit. The script sequentially implements the following functions:
# 1. Instantiate a UnifiedDataset by providing an accessible EEG dataset address.
# 2. Export a Locator that records metadata for dataset.
# 3. Instantiate a UnifiedDataset by providing a Locator.
# 4. Instantiate UnifiedDataset by integrating multiple UnifiedDataset instances.
# 5. Combine multiple datasets together.
# Example:
>>> from eegunity import UnifiedDataset, con_udatasets
# Approach (1) - Instantiate UnifiedDataset by providing an accessible EEG dataset address.
>>> unified_dataset = UnifiedDataset(domain_tag="dataset_demo", dataset_path='path/dataset')
>>> unified_dataset.save_locator('path/to/locator')
# Approach (2) - Instantiate UnifiedDataset by providing a Locator.
>>> unified_dataset = UnifiedDataset(domain_tag="dataset_demo", locator_path='path/locator')
# Approach (3) - Instantiate UnifiedDataset by integrating multiple UnifiedDataset instances.
>>> unified_dataset_1 = UnifiedDataset(domain_tag="dataset_1", locator_path='path/locator1')
>>> unified_dataset_2 = UnifiedDataset(domain_tag="dataset_2", dataset_path='path/dataset2')
>>> unified_dataset_list = [unified_dataset_1, unified_dataset_2]
>>> combined_datasets = con_udatasets(unified_dataset_list)
\end{lstlisting}
\vspace{-\baselineskip}
\rule{0.48\textwidth}{0.4pt}
\vspace{-\baselineskip}
\end{table}

\subsection{Data Correction}
\begin{figure}[htbp]
    \captionsetup{font={footnotesize}}
    \centering
    \includegraphics[width=1\linewidth]{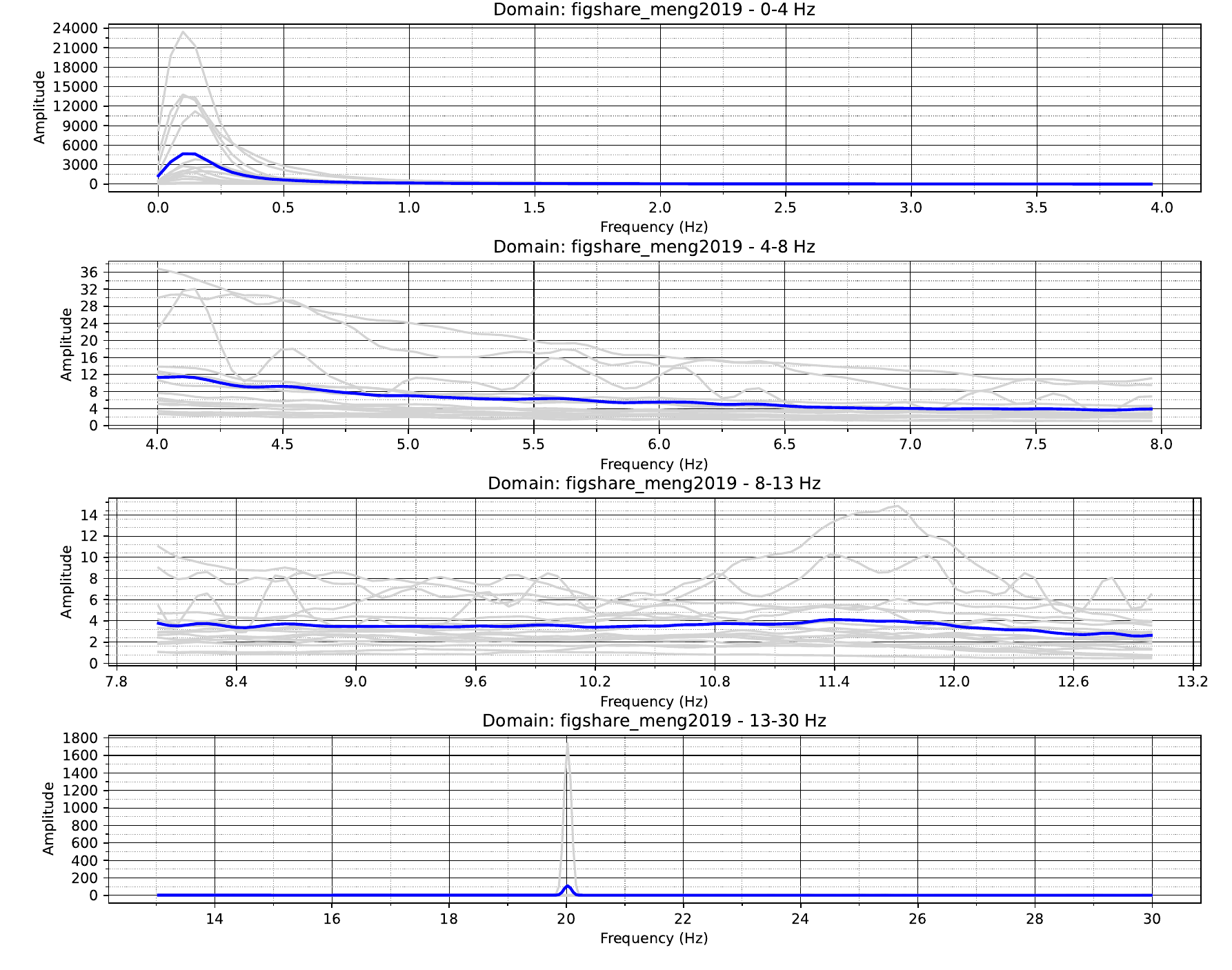}
    \caption{Frequency Visualization Results of Correction Module. The figure displays magnitude-frequency curves across four subfigures, with each subfigure corresponding to one of the frequency bands: alpha, beta, theta, and gamma. The samples visualized are randomly selected within a domain specified by a ``domain tag". The average curve for each band is represented in blue, while individual curves for each data sample are depicted in grey. \label{fig_vis_frequency}}
\end{figure}

\begin{figure}[htbp]
    \captionsetup{font={footnotesize}}
    \centering
    \includegraphics[width=1\linewidth]{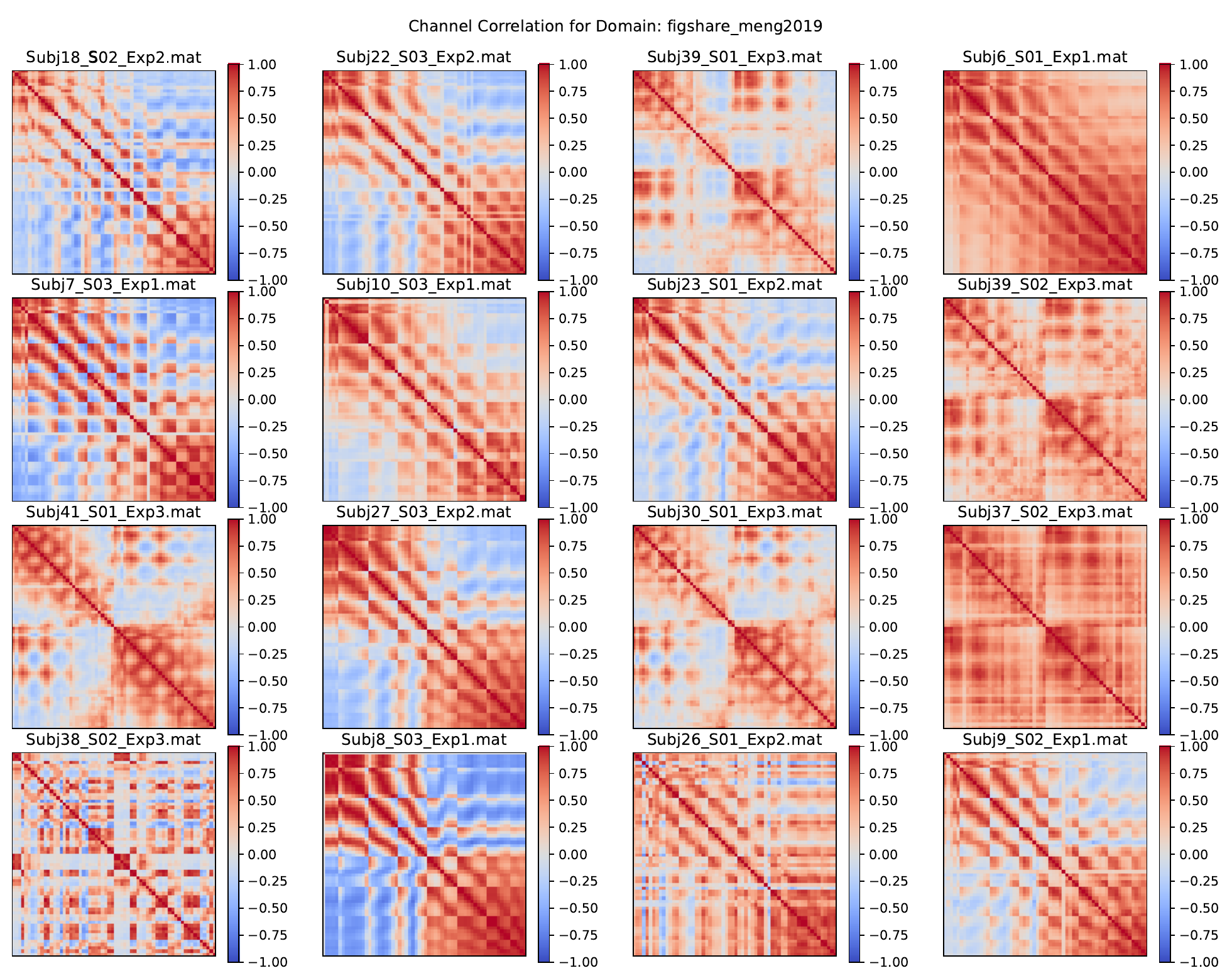}
    \caption{Channel Correlation Visualization Results of Correction Module. The figure presents channel correlation for samples that are randomly selected within a domain identified by a ``domain tag". The number of samples to be visualized is adjustable via a parameter. Subfigure placement is automatically optimized for ease of review. This figure allows users to inspect specific aspects of the samples, such as low-frequency noise or channel noise, facilitating a detailed analysis \cite{FilteringU0_2023_yan}. \label{fig_vis_channel_corr}}
\end{figure}

EEGUnity supports data correction for one or multiple EEG datasets, including a user-friendly interface for inspecting, modification, diagnosis, and visualization for dataset integrity. The implementation details of the aforementioned functions are outlined in Table \ref{code_correction} and elaborated below:
\begin{itemize}
    \item \textbf{Interface for reviewing and modification}: EEGUnity stores metadata in the \codesty{Locator}, which users can easily review and modify using tools like PyCharm, Microsoft Excel, and Pandas;
    \item \textbf{Dataset diagnosis}: EEGUnity supports dataset diagnosis through built-in functions that generate detailed dataset reports, including the ratio of file types, domain tags, channel configurations, sampling rates, and completeness checks;
    \item \textbf{Dataset visualization}: EEGUnity includes built-in functions to visualize datasets, such as displaying magnitude-frequency curves for alpha, beta, theta, and gamma waves and channel correlations for each data. Two visualization results are shown in Fig. \ref{fig_vis_frequency} and Fig. \ref{fig_vis_channel_corr}.
\end{itemize}

\begin{table}[htbp]
\captionsetup{font={footnotesize}}
\fontsize{8}{10}\selectfont
\centering
\scriptsize
\caption{Implementation for Diagnosing and Visualizing Datasets in EEGUnity. \label{code_correction}}
\rule{0.48\textwidth}{0.4pt}
\vspace{-\baselineskip}
\begin{lstlisting}[language=Python, basicstyle=\ttfamily]
# Diagnosing and Visualizing Datasets in EEGUnity
# -----------------------------------------------
# This script demonstrates how to diagnose and visualize EEG datasets using the EEGUnity toolkit. The script sequentially implements the following functions:
# 1. Instantiate a UnifiedDataset by providing an accessible EEG dataset address.
# 2. Generate and display a report summarizing EEG dataset.
# 3. Visualize the frequency spectrum of EEG signals for a subset of samples.
# 4. Visualize the correlation between EEG channels for a subset of samples.
# Example:
>>> from eegunity import UnifiedDataset
>>> unified_dataset = UnifiedDataset(domain_tag='dataset1', dataset_path='path/to/dataset')
>>> unified_dataset.eeg_correction.report()
>>> unified_dataset.eeg_correction.visualization_frequency(max_sample=10)
>>> unified_dataset.eeg_correction.visualization_channels_corr(max_sample=16)
\end{lstlisting}
\vspace{-\baselineskip}
\rule{0.48\textwidth}{0.4pt}
\vspace{-\baselineskip}
\end{table}

\begin{figure*}
    \captionsetup{font={footnotesize}}
    \centering
    \includegraphics[width=0.85\linewidth]{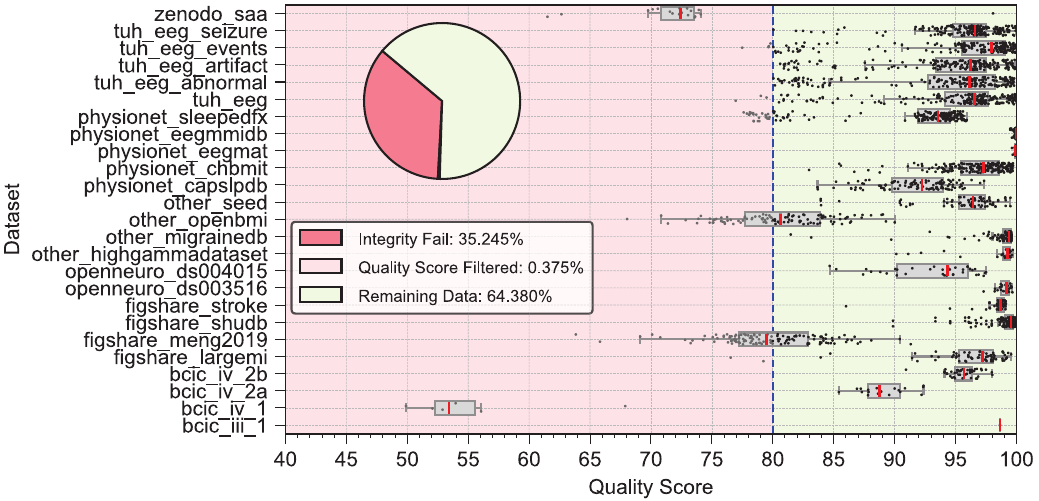}
    \caption{A Statistical Result of Data Integrity Checks and Quality Assessments for Filtering. The figure illustrates the results of sample filtering using two types of charts: a pie chart and a box plot. The pie chart displays the proportion of data filtered by completeness (i.e., fully completed data) and quality (i.e., data with scores above 80), along with the proportion of the remaining data. The pie chart provides an overview for users to understand the extent of data retained after filtering. The box plot shows the distribution of scores across various datasets. The red line within each box plot represents the median score, while individual dots around the box represent the scores of specific data points, with the maximum number of dots adjustable via parameters. These dots provide insight into the score distribution within each dataset. \label{fig_data_clean_result}}
\end{figure*}

\begin{table}[htbp]
\fontsize{8}{10}\selectfont
\captionsetup{font={footnotesize}}
\centering
\scriptsize
\caption{Implementation for Data Cleaning in EEGUnity. \label{code_data_cleaning}}
\rule{0.48\textwidth}{0.4pt}
\vspace{-\baselineskip}
\begin{lstlisting}[language=Python, basicstyle=\ttfamily]
# Data Cleaning Implementation in EEGUnity
# -----------------------------------------
# This script demonstrates the process of cleaning EEG data using the EEGUnity toolkit. The script sequentially implements the following functions:
# 1. Instantiate a UnifiedDataset object.
# 2. Filter EEG samples based on completeness.
# 3. Save the dataset state after filtering.
# 4. Filter sample based on quality assessment.
# 5. Apply a bandpass filter.
# 6. Perform Independent Component Analysis (ICA).
# 7. Save the final dataset state.
# Example:
>>> from eegunity import UnifiedDataset
>>> unified_dataset = UnifiedDataset(domain_tag="test", locator_path="./test_dataset")
>>> unified_dataset.eeg_batch.sample_filter(completeness_check="Completed")
>>> unified_dataset.save_locator("./test_dataset_completed")
>>> unified_dataset.eeg_batch.get_quality()
>>> locator = unified_dataset.get_locator()
>>> unified_dataset.set_locator([locator['Quality Score'] > 80])
>>> unified_dataset.eeg_batch.filter(output_path="./test_dataset_completed_bandpass_filter_1_49Hz", filter_type='bandpass', l_freq=1, h_freq=49)
>>> unified_dataset.save_locator("./test_dataset_completed_bandpass_filter_1_49Hz.csv")
>>> unified_dataset.eeg_batch.ica(output_path="./test_dataset_completed_bandpass_filter_1_49Hz_ica", max_components=20, method='fastica')
>>> unified_dataset.save_locator("./test_dataset_completed_bandpass_filter_1_49Hz_ica.csv")
\end{lstlisting}
\vspace{-\baselineskip}
\rule{0.48\textwidth}{0.4pt}
\vspace{-\baselineskip}
\end{table}

\subsection{Data Cleaning}
EEGUnity supports data cleaning for one or multiple EEG datasets, with the detailed implementation outlined in Table \ref{code_data_cleaning}. The suggested data cleaning process is divided into the following steps:
\begin{itemize}
    \item \textbf{Data completeness check}: EEGUnity classifies the completeness of EEG data into three levels and records the results in basic attributes of the \codesty{Locator}: (1) completed, indicating that the data includes the original EEG sequence and all basic attributes in the \codesty{Locator} have been correctly parsed; (2) acceptable, indicating that further parsing of EEG information may affect the accuracy of analyses; (3) unavailable: indicating the original EEG sequence is unavailable or any basic attributes in the \codesty{Locator} cannot be filled. Based on data completeness check, quality assessment, and sample filtering, a statistical result for multiple datasets is presented in Fig. \ref{fig_data_clean_result}.
    \item \textbf{Quality assessment}: EEGUnity is capable of supporting established methodologies \cite{quality_scores_Shady_2017} to comprehensively assess the quality of EEG data. The assessment results are presented as scores ranging from 0 to 100 and recorded in the advanced attributes of the \codesty{Locator} for future analysis.
    \item \textbf{Sample filtering}: EEGUnity supports filtering the data based on the attributes in the \codesty{Locator} to meet the specific requirements. The filtering process can take into account the quality and completeness level of the data, as well as custom criteria.
    \item \textbf{Denoising}: EEGUnity supports the use of independent component analysis and filtering techniques to remove noise \cite{NoiseU0_2024_li} from the EEG data \cite{MNEPython_Alexandre_2013}. The denoising process can significantly improve the quality of the data and ensure the reliability of subsequent analysis.
\end{itemize}

\begin{table}[htbp]
\captionsetup{font={footnotesize}}
\fontsize{8}{10}\selectfont
\centering
\scriptsize
\caption{Implementation for Data Unification in EEGUnity. \label{code_data_unification}}
\rule{0.48\textwidth}{0.4pt}
\vspace{-\baselineskip}
\begin{lstlisting}[language=Python, basicstyle=\ttfamily]
# Data Unification Implementation in EEGUnity
# --------------------------------------------
# This script demonstrates the process of unifying and preparing EEG data using the EEGUnity toolkit. The script sequentially implements the following functions:
# 1. Instantiate a UnifiedDataset object by specifying the path to the EEG dataset and a domain tag for identification.
# 2. Save the EEG data in a different format (e.g., FIF format).
# 3. Resample the EEG data to a new sampling frequency.
# 4. Align EEG channels to a specified order.
# 5. Infer the units of the EEG data.
# 6. Extract and display events from the EEG data and segment it into epochs.
# 7. Prepare segmented EEG data for pretraining.
# Example:
>>> from eegunity import UnifiedDataset
>>> unified_dataset = UnifiedDataset(dataset_path="./test_dataset", domain_tag="test_data")
>>> unified_dataset.eeg_batch.save_as_other(output_path="./test_dataset_fif")
>>> unified_dataset.eeg_batch.resample(output_path="./test_dataset_resample", new_sfreq=128)
>>> unified_dataset.eeg_batch.align_channel(output_path="./test_dataset_epoch", channel_order=["Cz", "C3", "C4"])
>>> unified_dataset.eeg_batch.infer_units()
>>> unified_dataset.eeg_batch.get_events()
>>> unified_dataset.eeg_batch.epoch_by_event(output_path="./test_dataset_epoch", seg_sec=1)
>>> unified_dataset.eeg_batch.epoch_for_pretraining(output_path="./test_dataset_epoch", seg_sec=1)
\end{lstlisting}
\vspace{-\baselineskip}
\rule{0.48\textwidth}{0.4pt}
\vspace{-\baselineskip}
\end{table}

\subsection{Data Unification}
Data unification is a novel batch processing workflow in EEGUnity, aiming to transform heterogeneous EEG data into a unified entity. The specific implementation of data unification varies according to different requirements and standards. A detailed outline of the implementation for data unification is illustrated in Table \ref{code_data_unification}. Currently, EEGUnity supports the following unification methods:

\begin{itemize}
    \item \textbf{Save as unified format}: EEGUnity supports saving EEG data in a specified unified format to facilitate subsequent processing and analysis.

    \item \textbf{Resampling}: EEGUnity supports resampling each EEG dataset to a specific target sampling rate.

    \item \textbf{Channel alignment}: EEGUnity supports aligning the channels of EEG data according to a specified channel order, where any missing channel is constructed using interpolation methods, ensuring completeness and consistency.

    \item \textbf{Normalization}: EEGUnity supports computing normalization transformation factors, including the variance and mean of samples/channels, and records the results in the \codesty{Locator}. The transformation factors are stored in the \codesty{Locator}, making it convenient for users to inspect and correct the factors. This normalization process can be performed when retrieving data.

    \item \textbf{Infer unit}: EEGUnity supports inferring the units for each channel in EEG data. The inferred units are stored in the \codesty{Locator}, enabling users to inspect and correct inferred units easily. Users can convert the units when retrieving data.

    \item \textbf{Extract event}: EEGUnity supports extracting events for each data and storing the results in the \codesty{Locator}. Users can quickly inspect, correct, and supplement the events in \codesty{Locator}. The events in \codesty{Locator} can be directly used with other epoch functions in EEGUnity.

    \item \textbf{Epoch by event}: EEGUnity supports extracting epochs (segments) from each EEG dataset based on specified events listed in the \codesty{Locator}.

    \item \textbf{Epoch for pretraining}: EEGUnity supports extracting epochs (segments) from each EEG dataset by controlling segment parameters.
\end{itemize}

\section{Discussion \label{section_discussion}}
EEG data processing capability has been significantly improved by software tools such as EEGLAB, FieldTrip, and MNE-Python. Initially, EEGLAB revolutionized EEG analysis by providing a user-friendly MATLAB-based GUI with advanced capabilities like independent component analysis. Following EEGLAB, FieldTrip offered a modular and flexible MATLAB-based toolbox that emphasized customized and in-depth analyses script-based approach. Subsequently, MNE-Python expanded the landscape by introducing Python packages for EEG data processing, with an emphasis on the reproducibility of data processing pipelines. The progression of established software tools reflects a trajectory toward increasingly sophisticated, flexible, and integrated EEG analysis methods.

Compared with established software, the novel Python package EEGUnity proposed in this paper provides notable advantages in multiple EEG datasets management and large-scale data processing, including various functions such as data correction, data cleaning, data unification, and customized batch operations. In addition to functionality, the flexibility of EEGUnity is another critical aspect: (1) the \codesty{Locator} design facilitates easy inspection and modification of dataset descriptions, ensuring accurate and reliable data processing; (2) the package features a user-friendly batch process interface, allowing researchers to develop customized batch processes tailored to specific requirements.

The ability to process large-scale data has become increasingly significant in the field of EEG, as foundation models require vast amounts of data for pretraining \cite{SelfSupervisedU0_2023_eldele}. This trend is similar to computer vision, where the availability of large and annotated datasets like ImageNet \cite{imagenet_deng_2009} have significantly boosted the development of foundational models. Similarly, the ability of EEGUnity to handle and integrate multiple datasets can drive significant advancements in EEG research, enabling the creation of powerful and accurate foundation models. The large foundational models are essential for extending advanced EEG tasks, eventually facilitating the application of BCI systems. Despite the advantages provided by EEGUnity, the construction of large-scale EEG datasets faces three challenges: data permission, privacy concerns, and data validation \cite{noisy_data_Khan_2023, privacy_preserving_Wang_2024}. Addressing these challenges requires coordinated efforts from the research community to establish guidelines and frameworks that facilitate data sharing while safeguarding individual privacy and ensuring data validation.

\section{Conclusion \label{section_conclusion}}
This paper introduces EEGUnity, a new Python package designed to manage multiple EEG datasets, representing a significant step forward in the management of large-scale EEG datasets. In the management of large-scale EEG datasets, EEGUnity provides powerful data processing capabilities, including data correction, data cleaning, and data unification, all of which can be applied flexibly through the \codesty{Locator} design and custom batch processing interface. The comprehensive capabilities and flexible design position EEGUnity as a valuable tool for researchers. The integration of large-scale, high-quality datasets facilitated by EEGUnity has the potential to help researchers gain a deep understanding of brain patterns and develop powerful EEG foundation models. As EEGUnity continues to evolve, the contribution of EEGUnity to efficiency and scalability will become pronounced. Future work includes addressing the challenges in data permission, privacy concerns, and data validation. Addressing these challenges requires coordinated efforts, potentially led by a representative organization that obtains necessary permissions and leverages EEGUnity to create a public, large, and standardized EEG dataset.

\bibliographystyle{plain}

\end{document}